\documentclass[11pt]{article}                  
\begin{document}

\def\det{{\rm det}}  \def\IR{{\bf R}}     \def\IL{{\bf L}} \def\map{T}
\def\la{\lambda}     \def\ep{\varepsilon} \def\bS{{\bf S}}
\def\n{\noindent} 
\def\CR{$$ $$}       \def\IP{{\bf P}}

\newtheorem{theorem}{Theorem}[section]       
section
\newtheorem{lemma}{Lemma}[section]           
section
\newtheorem{proposition}[lemma]{Proposition} 
\newtheorem{corollary}[lemma]{Corollary}     
\def\rmname{Remark}
\newtheorem{rmrk}[lemma]{\rmname}
\newenvironment{remark}{\begin{rmrk}\rm}{\end{rmrk}}     
\renewcommand\theequation{\thesection.\arabic{equation}}  
             \catcode`@=11 \@addtoreset{equation}{section} \catcode`@=12
\def\proof{\smallskip \noindent {\bf Proof. \ }}

\newcommand\filledsquare{\ \vrule width 1.5ex height 1.2ex} 
\def\qed{\hfill\filledsquare\linebreak\smallskip\par}

\title{Asymptotically exact spectral estimates \\
       for left triangular matrices}
\author{Michael Blank\thanks{This research has been partially
                             supported by INTAS, CRDF and RFBR grants.}
       \\ \\
        Russian Ac. of Sci., Inst. for
        Information Transmission Problems, \\
        B.Karetnij 19, 101447, Moscow, Russia, blank@iitp.ru}
\date{September 7, 2000}
\maketitle

\begin{abstract}
For a family of $n*n$ left triangular matrices with binary
entries we derive asymptotically exact (as $n\to\infty$)
representation for the complete eigenvalues-eigenvectors problem.
In particular we show that the dependence of all eigenvalues on
$n$ is asymptotically linear for large $n$. A similar result is
obtained for more general (with specially scaled entries) left
triangular matrices as well. As an application we study ergodic
properties of a family of chaotic maps.
\end{abstract}

\bigskip%
\n{\bf Keywords}: triangular matrix, spectrum, integral operator,
                  finite rank approximation, topological entropy

\bigskip%
\n{\bf AMS Subject Classification}: 
   Primary 37E15; Secondary 37D35, 37B15, 60K.

\section{Binary left triangular matrices}\label{section-bin-tr}
By an $n\times n$ left triangular matrix with binary entries we mean
the matrix $A_n=(a_{ij})$ whose entries up to (and including) the
secondary diagonal are ones, while all others are zeros, i.e.
$a_{ij}=1$ if and only if $1\le i \le n-j+1$. Despite its very
classical appearance spectral properties of such matrices were not
known, probably due to the fact that for a finite matrix size $n$
there is no reasonable representation for the spectrum. Denote by
$\la_1^{(n)}, \la_2^{(n)}, \dots, \la_n^{(n)}$ the eigenvalues of the
matrix $A_n$ ordered by their modules (in the decreasing order) and
by $\{e_k^{(n)}\}_{k=1}^n$ the corresponding orthogonal system of
eigenvectors (which exists due to symmetry of the matrix). As usual
by $(e)_i$ we denote the $i$-th entry of the vector $e\in\IR^{n}$ and
by $||e||$ -- its $\IL^2$-norm.

We start from numerical results. It turns out that numerically
with very high accuracy for each fixed $k\in\{1,\dots,n\}$ the
dependence of the k-th eigenvalue $\la_k^{(n)}$ is linear on $n$.
From this result one might expect that the determinant
$\det(A_n)$ grows as $n^n$ for large $n$, however expanding the
matrix with respect to the last line one shows that
$\det(A_n)=(-1)^{n+1}$. So there is no surprise that another
numerical test shows that for a fixed value $n$ the modules of
eigenvalues $|\la_k^{(n)}|$ decay hyperbolically on $k$. The
following statement gives an asymptotic representation for the
spectrum, which confirms these numerical findings.

\begin{theorem}\label{t=triang-bin}
Let $A_n$ be the left triangular matrix. Then for each fixed
$k\in\{1,\dots,n\}$
$$  \la_k^{(n)} = (-1)^{k+1} \frac{n}{(k-1/2)\pi} + O(1) ,$$
$$  (e_k^{(n)})_i
             = \cos\left(\frac{(k-1/2)(i-1)}{n}\pi \right)  + O(1/n) .$$
\end{theorem}

The {\bf proof} is based on the following statements. First notice that
$(A_n v)_i = \sum_{j=1}^{n-i+1} v_j$ for any vector $v\in\IR^n$.

\begin{lemma}\label{l=apr}
For any $C^1$ function $f:[0,1]\to [-1,1]$ with the derivative not
exceeding 1 by absolute value we have
$$ \sum_{j=1}^{n-i+1} f((j-1)/n)
  = n \sum_{j=1}^{n-i+1} f((j-1)/n) \cdot \frac1n
  = n \int_0^{1-\frac{i-1}n} f(s)~ds + O(1) .$$
\end{lemma}

\proof Direct computation. \qed

\begin{lemma}\label{l=eig}
The operator $L: f(x) \to \int_0^{1-x} f(s)~ds$ considered as an operator
acting in the space $\IL^2$ has the complete system of orthogonal
eigenfunctions
$$ E_k(x) := \cos((k-1/2)\pi x) $$
with eigenvalues
$$ \mu_k:= (-1)^{k+1} \frac{1}{(k-1/2)\pi} $$
for $k=1,2,\dots$.
\end{lemma}

\proof Indeed,
$$ \int_0^{1-x} \cos((k-1/2)\pi s)~ds
 = -\frac1{(k-1/2)\pi} \sin((k-1/2)\pi s)|_{0}^{1-x} \CR
 = \frac1{(k-1/2)\pi} \sin((k-1/2)\pi(1-x))
 = \frac{(-1)^{k+1}}{(k-1/2)\pi} \cos((k-1/2)\pi x) .$$
\qed

\begin{lemma}\label{l=matr-ineq}
Let the equality $Av = \mu v + \xi$ be satisfied for a symmetric
matrix $A$, two vectors $v,\xi\in\IR^n$ with $||\xi|| \le \ep||v||$
and a scalar $\mu$. Then the closest to $\mu$ eigenvalue $\la$ of the
matrix $A$ satisfies the inequality $|\la-\mu|\le\ep$ and if its
multiplicity is equal to 1, then the corresponding eigenvector $e$ is
such that $||e-v||\le O(\ep)\cdot||v||$.
\end{lemma}

\proof Probably this statement is not quite new (see for example
close statements in \cite{Pa}), however since we were not able to
find exact references to needed estimates we give a sketch of the
proof. Notice that $(A - \xi v^{*}/||v||^{2})v=\mu v$, i.e. the
vector $v$ is the eigenvector of the perturbed matrix $A_{\mu}
:= A - \xi v^{*}/||v||^{2}$. Indeed,
$$ A_{\mu}v = \mu v + \xi - \xi v^{*}v/||v||^{2}
            = \mu v + \xi - \xi = \mu v .$$
On the other hand,
$$ ||\xi v^{*}/||v||^{2}|| \le \frac{||\xi||}{||v||} \le \ep .$$
Thus the matrix $A_{\mu}$ can be considered as an
$\ep$-perturbation of the symmetric matrix $A$. Thus the desired
estimates can be obtained by the standard perturbation argument.

Moreover, to prove the first inequality about the closest
eigenvalue $\la$ one can use a more direct argument \cite{Pa}.
Namely, for $\mu=\la$ the inequality becomes trivial, while
otherwise
$$ ||v|| \le ||(A - \mu I)^{-1}|| \cdot ||(A - \mu I)v||
 = \frac1{|\mu-\la|}\cdot ||\xi|| .$$
Thus
$$ |\mu-\la| \le \frac{||\xi||}{||v||} \le \ep .$$
\qed

\n{\bf Proof} of Theorem~\ref{t=triang-bin}. It suffices to show
that for each $k=1,2,\dots,n$ a projection to the space of
piecewise constant functions of the $k$-th eigenvector of the
linear operator $L$ introduced in Lemma~\ref{l=eig} satisfies the
equality of type described in Lemma~{l=apr} with $\ep=O(1)$. Let
$\IP_{n}:C^{1}([0,1]\to\IR^{1}, R^{n})$ be a projection operator
acting from the space of $C^{1}$ functions $f:[0,1]\to\IR^{1}$ to
$\IR^{n}$ defined as follows:
$$ (\IP_{n}f)_{i} := f((i-1)/n) .$$
Then, according to Lemma~\ref{l=apr}, the following equality
holds true:
$$ A_{n} \IP_{n}E_{k} = n\mu_{k}\IP_{n}E_{k} + \xi $$
for a vector $\xi\in\IR^{n}$ such that $||\xi||\le O(1)$. Now the
application of Lemma~\ref{l=matr-ineq} finishes the proof. \qed

\section{Properly scaled left triangular matrices}\label{section-gen-tr}

Results of the previous section might be generalized for a more
general class of left triangular matrices. Let
$\phi:[0,1]\to(0,1]$ be a $C^{1}$ positive function. We define a
family of $n\times n$ left triangular matrices generated by the
function $\phi$ as follows:
$(A_{n}^{\phi})_{i,j}:=\phi((j-i+1)/n)$ for $i\le j$ and zero
otherwise. Clearly the binary left triangular matrices satisfy
this property for $\phi\equiv1$.

Again similarly to the argument in the previous section one can
prove the asymptotically linear dependence on $n$ of the
eigenvalues of the matrix $A_{n}^{\phi}$ using the property that
the main contribution to the ``shape'' of eigenvectors comes from
the corresponding eigenfunctions of the integral operator
$L_{\phi}:f(x) \to \int_{0}^{1-x}\phi(s)f(s)~ds$.

\begin{theorem}\label{t=gen-tr} Let $\mu$ and $f_{\mu}$ be
respectively an eigenvalue and an eigenfunction of the operator
$L_{\phi}$ (i.e. $L_{\phi}f_{\mu} = \mu f_{\mu}$). Then for each
positive integer $n$ there exists an eigenvalue $\la=n\mu+O(1)$
and an eigenvector $e\in\IR^{n}$ of the matrix $A_{n}^{\phi}$
such that $(e)_{i}=f_{\mu}((i-1)/n) + O(1/n)$ for each
$i=1,2,\dots,n$. \end{theorem}

Since in this more general case we are not able to give an
explicit expression for the asymptotic spectrum, we shall
demonstrate also another (more direct) way to derive the
asymptotic linearity of the spectrum. Assume that the function
$\phi$ is nonincreasing and denote by $S_1^n$ the set of $C^1$
monotonically decreasing concave positive functions
$f:[0,1]\to[0,1]$ with $f(0)=1$ and by $\bS_1^n\subset\IR^n$ --
the set of vectors $v$ such that $v_i:=f((i-1)/n)$ for some $f\in
S_1^n$. We introduce also the following nonlinear operator
$B_n:\IR^n\to\IR^n \cup (\infty)^n$ defined by the relation:
$(B_n v)_i := (A_n^{\phi}v)_i/(A_n^{\phi}v)_1$.

\begin{lemma}\label{l=space} The set $\bS_1^n$ is invariant with
respect to the operator $B_n$ and for any vector $v\in\bS_1^n$
the sequence $\{B_n^m v\}_m$ converges as $m\to\infty$ in
$\IL^2$-norm to the leading eigenvector $e_1^{(n)}$ of the matrix
$A_n^{\phi}$.
\end{lemma}

\proof Since the coordinates of the vector $v$ are positive we
get that for any $i$
$$ (A_n^{\phi}v)_{i+1} - (A_n^{\phi}v)_{i}
 = -\phi((n-i+1)/n)v_{n-i+1} < 0 .$$
On the other hand, due to the fact that the coordinates of the
vector $v$ decrease monotonically we obtain
$$ (A_n^{\phi}v)_{i+1} - 2(A_n^{\phi}v)_{i} + (A_n^{\phi}v)_{i-1}
 = \phi(\frac{n-i+2}n)v_{n-i+2} - \phi(\frac{n-i+1}n)v_{n-i+1} \CR
 = \phi(\frac{n-i+1}n)(v_{n-i+2} - v_{n-i+1})
 - \left(\phi(\frac{n-i+1}n) - \phi(\frac{n-i+2}n)v_{n-i+2} \right)
   v_{n-i+2} \CR
 < \phi(\frac{n-i+1}n)(v_{n-i+2} - v_{n-i+1}) < 0 .$$
These two inequalities prove the monotonicity and the concavity
of $A_n^{\phi}v$ respectively. The normalization by the positive
number $(A_n^{\phi}v)_1$ finishes the proof of the first
statement of Lemma~2.1. Now noticing that the matrix $A_n^{\phi}$
has only non negative entries we derive the second statement from
the well known Perron theorem. \qed

Thus choosing vectors $v$ from the set $\bS_1^n$ providing the
smallest and the largest contribution to $(A_{n}^{\phi}v)_{1}$
(namely $(\tilde v)_{i}=(n-i+1)/n$ and $(\hat v)_{i}\equiv1$
respectively) we immediately obtain that
$$ n/2 \int_{0}^{1}(1-s)\phi(s)~ds + O(1) \le \la_1^{(n)}
  \le n \int_{0}^{1}\phi(s)~ds + O(1).$$
In a similar (but somewhat involved) way one can study also other
eigenvalues and eigenvectors of this matrix.

\section{Application to chaotic dynamics}\label{section-appl}

Now we apply above results for the analysis of ergodic properties
of a certain family of chaotic maps. We refer the reader for the
necessary definitions and statements for example to \cite{ME,Bl}.

Fix a positive integer $n$ and consider a piecewise linear map
map $\map_n:[0,1]\to[0,1]$ from the unit interval into itself
defined as
$$ \map_n|_{[(i-1)/n,i/n)}(x):=\frac{i}{n}x $$
for all $i\in\{1,2,\dots,n\}$. This map is topologically mixing
and its second iterate is piecewise expanding (i.e. the modulus
of the derivative of the map is larger than $1$ for all points
where it is well defined), therefore this map has a unique
absolutely continuous invariant measure (see \cite{Bl}).

\begin{lemma} The topological entropy of the map $\map_n$
satisfies the equality
$h_{{\rm top}}(\map_n) = \ln\left(\frac{2n}{\pi} + O(1)\right)$.
\end{lemma}

\proof Notice that the map $\map_n$ is Markov with respect to the
partition to its intervals of monotonicity, i.e. it maps each interval 
of monotonicity to the union of some other intervals of monotonicity. 
Therefore one can construct the corresponding topological dynamics 
\cite{ME} as a shift operator $\sigma_n$ in the state of sequences 
$\bar y=\{y_{1},y_{2},\dots\}$ with the alphabet consisting of $n$ 
symbols (i.e. $y_{i}\in\{1,2,\dots,n\}$ and 
$(\sigma_{n}\bar y)_{j}:=y_{j+1}$), satisfying the property that for 
each $k\in\{1,2,\dots,n\}$ the symbol $k$ may be followed by any symbol 
from the set $\{1,2,\dots,n-k+1\}$, but not the symbol from the 
complement to this set. Thus the corresponding transition matrix $A_n$ 
\cite{ME} (matrix consisting of zeros and ones, describing possible 
transitions between symbols of the alphabet under the dynamics) is the 
left triangular binary $n\times n$ matrix. Now since the topological 
entropy $h_{{\rm top}}$ for both the Markov map $\map_n$ and the shift 
operator $\sigma_n$ is known to be equal to the logarithm of the largest 
eigenvalue of the transition matrix \cite{ME} we derive from 
Theorem~\ref{t=triang-bin} the desired relation for the topological 
entropy. \qed

\end{document}